# Ultra-High Lithium Storage Capacity of Al$_2$C Monolayer under Restricted Multilayered Growth Mechanism

Ning Lu,[a,*,1] Kai Wang,[a,1] Jiaxin Jiang,[a] Hongyan Guo,[a] Gui Zhong Zuo,[b] Zhiwen Zhuo,[a,*] Xiaojun Wu,[c] & Xiao Cheng Zeng[d,*]

[a] *Anhui Province Key Laboratory of Optoelectric Materials Science and Technology, Key Laboratory of Functional Molecular Solids Ministry of Education, and Department of Physics, Anhui Normal University, Wuhu, Anhui 241000, China*

[b] *Institute of plasma physics, HIPS, Chinese academy of Sciences, Hefei, 230031, China*

[c] *Hefei National Laboratory for Physical Sciences at the Microscale, CAS Key Laboratory of Materials for Energy Conversion, and School of Chemistry and Materials Sciences, University of Science and Technology of China, Hefei, Anhui 230026, China*

[d] *Department of Chemistry, University of Nebraska Lincoln, Lincoln, NE 68588, USA*

*Correspondence and requests for materials should be addressed to N.L. (email: luning@shu.edu.cn) or to Z.Z. (email: zhuozw@ustc.edu.cn) or to X.C.Z. (email: xzeng1@unl.edu).

[1] These authors contributed equally: Ning Lu, Kai Wang

**Designing anode materials with high lithium specific capacity is crucial to the development of high energy-density lithium ion batteries. Herein, a distinctive lithium growth mechanism, namely, the restricted multilayered growth for lithium, and a strategy for lithium storage are proposed to achieve the balance between the ultra-high specific capacity and the need to avert uncontrolled dendritic growth of lithium. In particular, based on first-principles computation, we show that the Al$_2$C monolayer with planar tetracoordinate carbon structure can be an ideal platform for realizing the restricted multilayered growth mechanism as a 2D anode material. Furthermore, the Al$_2$C monolayer exhibits ultra-high specific capacity of lithium of 4059 mAh/g, yet with a low diffusion barrier of 0.039-0.17 eV as well as low open circuit voltage in the range of 0.002-0.34 V. These novel properties endow the Al$_2$C monolayer a promising anode material for future lithium ion batteries. Our study offers a new way to design promising 2D anode materials with high specific capacity, fast lithium-ion diffusion, and safe lithium storage mechanism.**

The growing consumption of fossil fuels and the urgent need to address the global warming issue have led to the rapid exploit and development of clean and renewable energy sources.[1-5] As a leading technology for the storage of renewable energy, lithium-ion batteries (LIBs) have been widely used in every aspect of people's daily life owing to their high energy density, fast charging speed, long cycle life with few issues of environmental pollution.[6-9] As the host for lithium storage, various anode materials for LIBs have been developed with diverse chemical designs. Common ways involved in charging the host structure include lithium growth, adsorption, insertion, and lithium-host recombination, which could lead to varying degrees of destruction of the host structure. For the lithium-growth way, metallic lithium, an ideal anode material with ultra-high specific capacity, is commonly embraced. However, the propensity towards dendritic growth on the anode during the charging has become a major safety concern for practical application.[10] For the lithium-insertion way, graphite has been widely used as the anode material in LIBs mainly owing to its high electrical conductivity, high stability and accessibility.[11] However, the storage specific capacity ($C_M$) of graphite is merely 372 mAh/g, much lower than $C_M$ values of the state-of-the-art anode materials.[12,13] For the lithium-host-recombination way, silicon has been considered as a promising next-generation anode material due largely to its ultra-high theoretical specific capacity of up to 4200 mAh/g (via forming the $Li_{22}Si_5$ alloy).[14,15] Moreover, the highest alloyed state ($Li_{15}Si_4$) at room temperature holds a high specific capacity of 3579 mAh/g.[16,17] However, the excessive volume expansion of silicon anode (close to 300%) in the charging process can result in rapid decline of its recyclable capacity.[18] Although the volume-expansion problem can be resolved by making nano or porous structured Si,[19,20] other issues associated with Si anode such as the poor conductivity and low lithium ion diffusion are also need to be resolved.[21,22] Hence, anode materials with ultra-high specific capacity while free of the issues stated above are still under intense research and development.[23-27]

The two-dimensional (2D) materials can be an ideal platform for exploiting the lithium-adsorption way as anode materials, owing to many of their intrinsic and desirable characteristics, such as large specific surface area, abundant active sites, advantageous interlayer space, good thermal stability and unique electronic properties.[28] To date, many 2D anode materials with improved properties, including higher specific capacity and electrical conductivity, as well as lower energy barrier ($E_b$) for lithium-ion diffusion, have been predicted or realized in the laboratory. For instance, the theoretical specific capacity of graphene can reach up to twice the capacity of the graphite,[29] while the experimental reversible storage capacity amounts to 794-1054 mAh/g.[30] Phosphorene has been reported to have low energy barrier of 0.08 eV for lithium-ion diffusion (in certain crystalline directions) and a specific capacity of 432mAh/g.[31,32] Borophene has been predicted to have an ultra-low barrier (0.0026 eV) for lithium-ion diffusion and very high specific capacity of 1860 mAh/g.[33] Notably, some 2D Mxene materials, with transition-metal-atom-covered surfaces, entail high multilayered lithium-adsorption capacity and low barrier for lithium-ion diffusion. For example, V$_2$C exhibits two layered adsorption and E$_b$ of 0.045 eV,[34] and Ti$_2$C$_3$ exhibits one layered adsorption and E$_b$ of 0.07 eV.[35] Although many 2D anode materials exhibit lower energy barrier for lithium-ion diffusion than the graphite (E$_b$ = 0.45 eV[36,37]) and silicon (E$_b$ = 0.58 eV[38]), their specific capacities are still much lower than that of silicon or metallic lithium. Today, 2D anode materials with ultrahigh specific capacity comparable to silicon or metallic lithium are still highly sought.

In this work, we report theoretical evidence that 2D Al$_2$C monolayer, with a planar tetracoordinate carbon (ptC) structure ,[39,40] can be a highly promising anode material for LIBs. Our first-principles computation indicates that the Al$_2$C monolayer entails an ultrahigh lithium storage specific capacity of 4059 mAh/g under the restricted multilayer growth mechanism, and a low energy barrier of lithium-ion diffusion (E$_b$ = 0.039 eV). The two properties rival that of 2D Mxene and silicon, respectively.

## Results

**Structural and electronic properties of Al$_2$C monolayer.** The Al$_2$C monolayer is a planar structure with group symmetry of $D_{2h}$ (space group: No.47) and its unit cell consists of 2Al and 1C, as shown in Fig. 1a. The optimized lattice constants are $a$ = 3.04 Å and $b$ = 5.06 Å, respectively. Each C atom in the monolayer is bonded with four adjacent Al atoms to form a ptC structure. In each ptC structure, the length of the Al-C bond is 1.96 Å, and the distance between the two Al atoms in the **b** direction is 2.47 Å. Two adjacent ptC structures are connected in the **b** direction via Al-Al bonds with a bond length of 2.60 Å. As shown in Fig. 1b, the computed electronic band structure suggests that the Al$_2$C monolayer is a semiconductor with an indirect band gap of 0.45 eV. The computed density of states (DOS) indicates that electronic states of the structure is mainly contributed by the p orbital of C atom and Al atom. All computational results are consistent with the previous publications.[39,40]

**Diffusion of Li atom on surface of Al$_2$C monolayer.** To assess the diffusion behavior of Li atom on Al$_2$C monolayer as an anode material, various adsorption sites and diffusion paths of a single Li atom are examined. As shown in Fig. 1c, four stable adsorption sites, labeled as T, H, B$_1$ and B$_2$, on the Al$_2$C monolayer are identified, whose corresponding adsorption energy is -1.629, -1.464, -1.603 and -1.081 eV, respectively. The T site appears to be the most favorable adsorption site. The Bader charge analysis indicates a charge transfer of 0.88 e from Li to Al$_2$C at the T site. This charge transfer validates the interaction between Li and C (see Supplementary Fig. 1 and Supplementary Table 1). Considering the stable adsorption sites and structural symmetry, three possible diffusion paths for adsorbed Li atom, between the nearest adjacent adsorption sites are examined, including path 1 (P1: T→B$_1$→T), path 2 (P2: T→H→T) and path 3 (P3: H→B$_2$→H), as shown in Fig. 1c. We used the CI-NEB method to calculate the corresponding energy barrier of lithium diffusion along the three paths, which is 0.039, 0.165 and 0.381 eV, respectively. Among the three paths, P1 (in direction **a**) and P2 (in direction **b**) appear to be the more favorable ones across the entire 2D structure, especially for P1. Note that the energy barrier of 0.039 eV for P1 is much lower than that of conventional anode material, e.g., graphite (0.45 eV[36,37]), silicon (0.58 eV[38]) and graphene (0.32 eV[41]), even slightly lower than that of phosphorene (0.08 eV[31]), and comparable to those of MXenes (0.068 eV for Ti$_3$C$_2$[35], 0.045 eV for V$_2$C[34] and 0.024 eV for Mn$_2$C[42]). In sum, our calculation results suggest fast diffusion of lithium atoms on the Al$_2$C monolayer, another important factor for designing desired 2D anode materials.

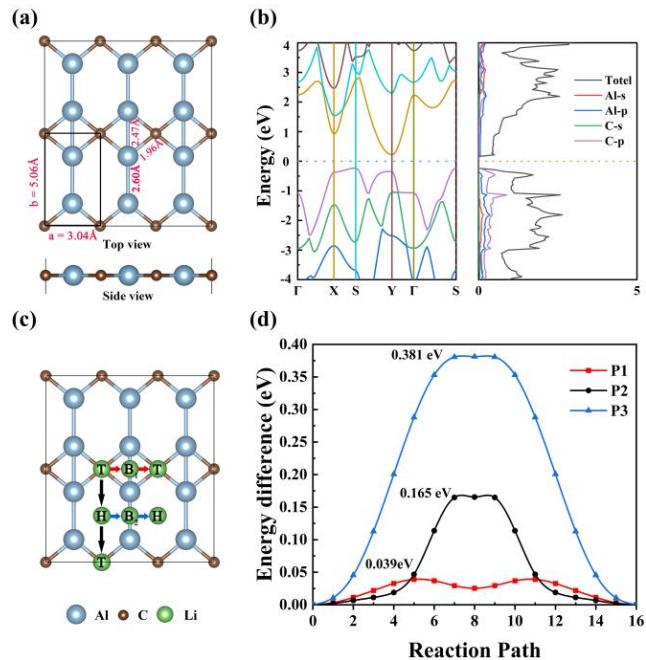

**Fig. 1.** (a) The top view and side view of the optimized Al$_2$C monolayer 3 × 2 × 1 supercell. (b) Computed electronic band structure and DOS of the Al$_2$C monolayer. (c) Adsorption sites and diffusion paths of Li on Al$_2$C monolayer. Here, T, H, B$_1$ and B$_2$ represent top site of C, hollow site of 4Al$_2$C ring, bridge site "1" between two Al atoms, and bridge site "2" between two Al atoms, respectively. (d) The calculated energy barrier for lithium diffusion along three most likely paths of P1, P2 and P3.

**A desirable lithium storage mode: Restricted multilayered slab of lithium on 2D anode materials.** As shown in Fig. 2, different lithium storage mode on 2D anode materials can be viewed as different stage of metal formation/growth process on a surface. If the attachment of lithium on the surface is through a layer-by-layer growth process, three stages can be identified in general: Distributed adsorption stage → multilayered slab stage → unlimited growth stage. The lithium storage mode corresponding to the distributed adsorption stage is of high safety but possesses low specific capacity. On the other hand, the lithium storage mode corresponding to the unlimited growth stage possesses extremely high specific capacity but suffers very low safety. In the intermediate multilayered slab stage, both good safety and ultra-high specific capacity can be achieved, thereby becoming a desirable lithium storage mode, especially when the thickness of the slab could be controlled by the metal-substrate interaction or other means.

On 2D anode materials, it is expected that the lithium slab can form on both sides of the 2D sheets so that the specific capacity can be doubled. Moreover, it is expected that 2D materials with the thinner structure and lighter atoms would enable thicker thickness of the intermediate multilayered slab, thereby higher theoretical specific capacity. The 2D Al$_2$C monolayer appears to be such an ideal candidate as a 2D anode material due to its single-atom-thickness, constituent of light atoms Al and C, and adaptation of restricted multilayered growth (see below for details).

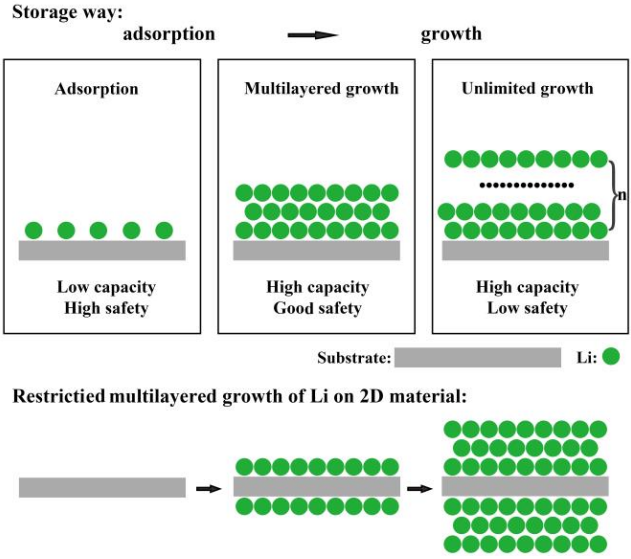

**Fig. 2.** Illustration of three different storage modes, including distributed adsorption (upper left panel), multilayered slab (upper middle panel), and unlimited growth on a substrate (upper right panel), with different levels of the specific capacity and safety concern. On 2D anode materials, the multilayered growth of Li metal slab can proceed on both surfaces (lower panel).

Lithium ion storage capacity under restricted multilayered growth mechanism, open circuit voltage and other related properties. The amount of lithium adsorbed and grown on the anode material determines the charge storage capacity of the LIBs. The half-cell reaction on Al$_2$C monolayer in relation to Li and Li$^+$ can be given as follows:

$$Al_2C + xLi^+ + xe^- \leftrightarrow Li_xAl_2C$$

Here, the 3×2×1 supercell is still used for simulation. Based on the single Li atom adsorption on $Al_2C$ illustrated above, our further examination of additional Li attachment onto $Al_2C$ monolayer can be considered in two stages: adsorption stage and growth stage. Specifically, the adsorption stage is defined such that only one layer of Li (n=1) is attached to each side of $Al_2C$, while the growth stage is considered to be layer-by-layer sequential attachment of Li up to five layers (n=2-5), as shown in Fig. 3a. For the first layer (n=1) of Li adsorbed on $Al_2C$, every two Li atoms tend to bond with a C atom on both up and down sides of $Al_2C$, forming a body-centered octahedron C-$Al_4Li_2$ (as shown in Supplementary Fig. 2). Upon Li adsorption, computed electron localization function (ELF), charge differential charge density, bond length, and Bader charges all indicate that there is strong ionic interaction between C-Li, which results in notable adsorption energy of -1.945 eV (see more details in Supplementary Information).

Upon further layer-by-layer growth of Li up to five layers (n=2-5), the adsorption is gradually weakened, as the adsorption energy changes from -1.647 eV to -1.607 eV. Importantly, for n = 6, the adsorption energy becomes -1.593 eV, whose absolute value is less than that of the critical value (-1.605 eV, vs. Li bulk cohesive energy), suggesting that the growth arrest of lithium multilayer slab would occur at n = 6. We name this growth arrest behavior as restricted multilayered growth. As such, the saturated adsorption capacity of the $Al_2C$ monolayer within the 3 × 2 × 1 supercell would be 60 Li atoms, with five lithium layers on each side of the $Al_2C$ monolayer. Thus, the stoichiometric ratio under this saturation condition would be $Li_{10}Al_2C$, which, remarkably, corresponds to the maximum specific capacity of 4059.4 mAh/g. This specific capacity is much higher than that of other 2D materials, e.g., graphene (794-1054 mAh/g), silicene (954 mAh/g[43]), phosphorene (432 mAh/g[31]), $V_2C$ (941 mAh/g[34]) and $Mn_2C$ (887.6 mAh/g[42]), while it is comparable to that of silicon (e.g., 4200 mAh/g at $Li_{4.4}Si$).

Next, the thermal stability of the lithium multilayer slab attached to $Al_2C$ is examined by performing an AIMD simulation for which the temperature is controlled at 300 K. As shown in Supplementary Fig. 3, the adsorbed Li layers are largely intact, suggesting high possibility of the formation of the multilayered slab on the $Al_2C$ monolayer. To further explore the most thermodynamically stable structure with different adsorption ratios of $Al_2C$ monolayer, the variation of convex hulls with adsorption concentration is plotted in Fig. 3b for n = 1-5. The convex hull of the calculated $E_f$ versus Li content, $x$, is shown in Fig. 3d. Note that since all the calculation points are located on the convex hull, one to five layer lithium-adsorbed structures are all thermodynamically stable based on the phase thermodynamically stability criterion.[44] From the most stable structures at different adsorption concentrations as shown in the convex hull diagram, the theoretical discharge voltage profile is calculated (see Fig. 3c). It can be seen that the electrode voltage decreases from 0.34 V to 0.002 V for the adsorbed Li layers at different concentrations. In addition, the computed DOS, corresponding to one to five layers of Li attached to the Al$_2$C monolayer, is shown in Fig. 3d, respectively. The appearance of electronic states around the Fermi level is clearly seen in all cases, indicating that all systems with different degrees of lithiation are metallic. Moreover, the density of electronic states around the Fermi level increases with increasing the lithium content, suggesting good conductivity during the charging and discharging processes.

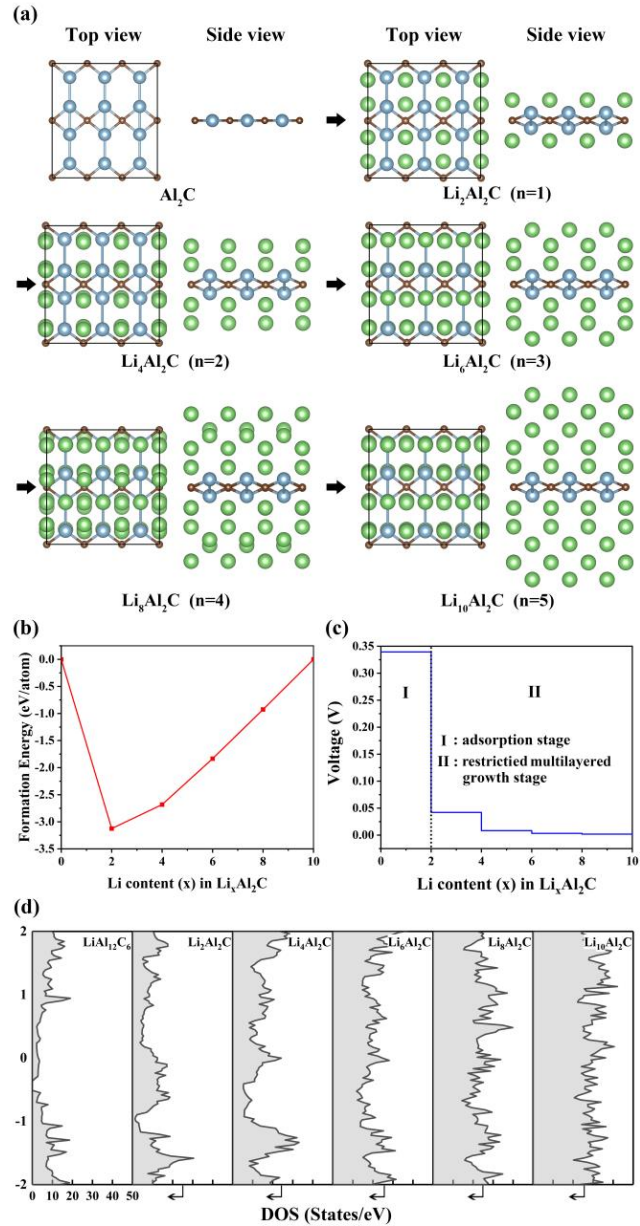

**Fig. 3.** (a) Layer-by-layer sequential attachment of Li on Al$_2$C monolayer. (b) Convex hulls of the formation energy at different concentration of Li. (c) Voltage versus concentration curve corresponding to the convex hulls. (d) Computed DOS corresponding to various adsorption ratios.

**Discussion**

To achieve ultra-high storage specific capacity of lithium of Al$_2$C monolayer, it is necessary to realize continuous attachment of lithium layer on both sides of the monolayer. For the first layer of the lithium, the good interaction between Li and C as discussed above and the enviable arrangement of Al atoms lead to easy lithium attachment at the adsorption stage, which supplies a favorable platform for further lithium attachment. To gain deeper insights into the restricted multilayered growth mechanism, various other properties on atomic Li arrangement have been analyzed (see below), including ELFs, charge transfer, work function, and lattice change (interlayer strain).

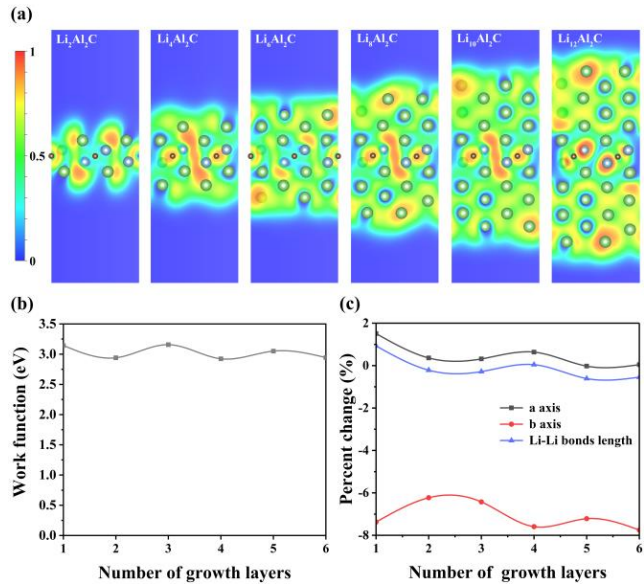

**Fig. 4.** (a) Computed ELFs of Al$_2$C monolayer with 1-6 layers of lithium attachment. (b) The work function versus the number of growth layers (n = 1-6). (c) Percentage change in lattice constant of Al$_2$C monolayer and in the surface Li-Li bond length versus the number of growing layers (n = 1-6). (More details see SI.)

As shown in Fig. 4a, the computed ELFs of Al$_2$C with different number of layers of lithium attachment show that electrons tend to disperse over each lithium slab with some electrons partially localized around Al$_2$C monolayer. Especially, there are intense electron density on the surfaces for all structures within n=1-6, thereby allowing the increasing number of lithium layers to grow. The charge transfer from the lithium slab to Al$_2$C monolayer are listed in Supplementary Table 2. Although the charge transfer exhibits significant changes between the first layer and the next two lithium layers during the attachment, it remains around 2.5 $e$ per Al$_2$C formula for larger n. Specifically, for n = 1, the strong Li-C interaction leads to notable charge transfer of 0.82 $e$ in average but only 1.64 $e$ per unit cell due to the few number Li atoms. For n = 2, the increased amount of Li results in the maximal charge transfer of 3.09 $e$ in total. For n = 3, the charge localized between Li and Al for n=2 (as shown in Fig. 4a) are rebalanced by the newly attached lithium layer. For n = 4-6, the charge transfer from the lithium slab to Al$_2$C monolayer changes slightly, reflecting the continuous growth of the new layer. Further examination of the calculated work function of the system at multilayered growth stage exhibits alternative (odd-even) but slight changes within 0.2 eV from n = 1-6, as shown in Fig. 4b. In summary, the ELF, charge transfer, and the work function are the factors that favor the continuous growth of the lithium layer to achieve increasing specific capacity, but not the factors that restrict the continuous growth.

The percentage change in the lattice constants of the system at multilayered growth stage allow us to understand the gradually weakened Li adsorption energy from n = 1 to 6. The lattice constants in both direction **a** and **b** exhibit gradual oscillatory behavior, corresponding to gradual contraction and eventual stabilization of lithium slab growth, as shown in Fig. 4c and Supplementary Table 2. The trend in the lattice constant in direction **a** appears to be closely correlated with the trend in the percentage change of averaged Li-Li bond length. Ultimately, the averaged Li-Li bond length would converge for n=∞. In other words, as n increases, the lattice feature of bulk lithium (See more details in SI) would become increasingly dominate for the (Li slab)-Al$_2$C-(Li slab) system. The lattice mismatch between Al$_2$C monolayer and lithium slabs causes gradually weakened Li adsorption with increasing the lithium layer number. Eventually, when the absolute value of adsorption energy is less than that of the critical value, the multilayered growth stage ceases as $n_{max}$=5, suggesting the lattice mismatch between Al$_2$C monolayer and Li multilayered slabs is the key factor for restricting the continuous vertical growth of the lithium slab.

In conclusion, the structural, electronic properties and electrochemical properties of Al$_2$C monolayer as an anode material for LIBs are investigated based on first-principles calculations. The lithium attachment on the Al$_2$C monolayer exhibits two stages: adsorption stage and restricted multilayered growth stage. In the first stage, the lithium atom prefers to adsorb on both sites of C atom, while in the second stage, layer-by-layer proceeds until n = 5. The lattice mismatch between Al$_2$C monolayer and lithium slabs causes gradually weakened Li adsorption with increasing the lithium layer number, resulting in the restricted multilayer growth mechanism. With the unique restricted multilayered growth mechanism on the lithium storage, the specific capacity of Al$_2$C monolayer amounts to 4059.4 mAh/g, rivaling that of silicon and metallic lithium. Additionally, the calculated lowest energy barrier for Li diffusion on the Al$_2$C monolayer is only 0.039 eV, suggesting high Li mobility for fast charge and discharge. The computed DOS indicates that the Al$_2$C monolayer attached with 1-5 layers of Li on each side all present metallic characteristics, thus ensuring the good electric conductivity as an anode material. Lastly, the compute open circuit voltage is in a desirable range of 0.002-0.34 V. Overall, this comprehensive study shows that the Al$_2$C monolayer is a promising 2D anode material for LIBs. Given its novel properties, it is our hope that the Al$_2$C monolayer based 2D anode material can be realized in the laboratory in near future. Our study also offers a new mechanism, the restricted multilayered growth mechanism, to design and develop promising 2D anode materials with high specific capacity, fast lithium-ion diffusion, and safe lithium storage.

## Methods

All the density functional theory calculations are carried out using VASP 5.4 software package.[45,46] The Perdew-Burke-Ernzerhof (PBE)[47] method of generalized gradient approximation (GGA)[48] is used to describe electron exchange and correlation functional, while projector augmented wave (PAW)[49] method is adopted to treat electron-ion interactions. With the plane-wave basis set, the cut-off energy is set to 500 eV (which is sufficient for energy convergence). All structures are fully optimized until the residual forces on all atoms are less than 0.01 eV/Å and total energy convergence of 10$^{-5}$ eV is achieved. The vacuum space is set at 16 Å so that the interaction between images of layer can be neglected for the calculation of the electronic properties of the Al$_2$C monolayer and the search for the most stable adsorption site. For computing the saturated adsorption capacity, the maximum vacuum layer increases to 30 Å. The k-point grid is dense enough to converge energy and geometry. The Bader charge is computed to analyze the charge transfer between lithium ion and Al$_2$C monolayer.[50] By using the climbing image nudged elastic band (CI-NEB) method,[41] the energy barrier for lithium diffusion on seven fully relaxed structures is calculated in order to study the kinetic mechanism of lithium diffusion on the Al$_2$C monolayer. Lastly, the thermal stability of structures of Al$_2$C attached with lithium multilayers is examined by using ab initio molecular dynamics (AIMD) simulation for 5 ps (per system) with a time step of 1 fs.

**Data availability.** All relevant data are available from the authors

**Acknowledgement**

NL was supported by the Anhui Provincial Natural Science Foundation (No.2008085QA33) and the National Natural Science Foundation of China (No. 11775261).

**Supplementary Information**. Calculation formulas of adsorption energy, stepwise adsorption energy, theoretical capacity and formation energy; the optimized structure of each adsorption site and the electronic properties of site T after adsorption of Li; the adsorption energy, the amount of charge transfer and the distance between Li and some atoms at each adsorption site; voltage profile for Li growth in layers one to six; stepwise adsorption energy, charge transfer from lithium slab to $Al_2C$, lattice change, superficial Li-Li bond length, and work function for multilayer Li attachment; snapshots from the AIMD simulation at the temperature of 300 K of geometrical structures for the $Al_2C$ monolayer and one to five layers of Li after attachment at 5ps with a time step of 1fs; formation of octahedron C-$Al_4Li_2$.

**Competing financial interests:** The authors declare no competing financial interest.